\documentclass[prl,floatfix,twocolumn]{revtex4-2}
\usepackage{amsmath}
\usepackage{amsfonts}
\usepackage{mathrsfs}
\usepackage{epsfig}
\usepackage{graphicx}
\usepackage{dcolumn}
\usepackage{hyperref}
\usepackage{xcolor}
\usepackage{placeins}
\usepackage{subfigure}
\usepackage[hyphenbreaks]{breakurl}

\begin{document}
\title{Lateral inhibition in relaxation oscillators provides a basis for computation}
%A dynamical framework for computational logic in non-equilibrium steady
%states arising in oscillatory media} 
%Relaxation oscillators interacting via lateral inhibition encode a computational logic
%Lateral inhibition in relaxation oscillators yields a computational logic
%Lateral inhibition in relaxation oscillators provides a basis for computation
%coupled relaxation oscillators}
\author{A. Parveena Shamim$^{1}$, Shakti N. Menon$^1$ and
Sitabhra Sinha$^{1,2}$}
\affiliation{$^1$The Institute of Mathematical Sciences, CIT Campus,
Taramani, Chennai 600113, India.\\
$^2$Homi Bhabha National Institute, Anushaktinagar, Mumbai 400094,
India.
}
\date{\today}
\begin{abstract}
Coupled relaxation oscillators, realized via chemical or other means, can exhibit a multiplicity of steady states, characterized by spatial patterns resulting from lateral inhibition.
We show that perturbation-initiated transformations between these configurations, mapped to binary strings via coarse-graining, provide a basis for computation.
The rules governing these transitions emerge from an underlying effective energy landscape shaped by the global and local stabilities of these states. Our results suggest a framework by which far-from-equilibrium systems may encode a computational logic.
\end{abstract}
%\pacs{}
%\keywords{}

\maketitle
%\section{Introduction}
Complex spatio-temporal patterns seen in natural phenomena, ranging from the
intricate symmetric geometry of snowflakes~\cite{Langer1980} to the
%emergence of form in multicellular organisms over the course of development~\cite{Kondo}
development of form in multicellular organisms~\cite{Koch1994,Kondo2010,Lander2011}
and large-scale plankton patchiness in oceanic algal blooms~\cite{Gower1980,Abraham2000,Neufeld2001}, are associated
with processes that are typically far from equilibrium~\cite{Cross1993,Gollub1999}.
The behavior of the dissipative systems where such patterns arise
%The dissipative processes that underlie such patterns 
can be described using a dynamical framework, yielding
solutions ranging from waves, breathers, and solitons, to chaos~\cite{Hagberg1996,Haim1996}.
Each of the collective states, which are characterized by distinct emergent phenomena, encodes 
information about the instantaneous configuration of the system. 
Consequently, a transition from one pattern to another
can be viewed as a ``computation'' that results in the transformation of such information~\cite{Wolfram1985,Crutchfield2012,LasCuevas2016}. %include Crutchfield references
Indeed, the dynamical evolution of systems such as spin glasses, which flow down rugged energy landscapes to 
%a local minimum, 
one of many possible local minima~\cite{Gu2012},
have been used to implement neural network-inspired computation~\cite{Hopfield1986,Becker2020}. This paradigm has been successfully used to 
model brain function~\cite{Amit1989}
as well as to provide innovative solutions to hard combinatorial optimization problems~\cite{Kirkpatrick1983,Fu1986,Monasson1999}. 
However, it has been
challenging to formulate such a language, which connects dynamics and computation, 
for describing systems that are far from equilibrium.

%The dynamical evolution of physical systems can be viewed as 
%computation~\cite{Wolfram1985}.
%Dynamics and computational complexity can be viewed as physical 
%processes (cite Crutchfield). Several dynamical systems can be viewed 
%as implementing computation from an initial to a final state (cite 
%paper on cellular automata~\cite{Wolfram1983,Wolfram1984}). 
%Looking at this from the point of view of 
%statistical physics, spin glasses provide for equilibrium (albeit non-ergodic) 
%[cite Hopfield, Physics Today] systems a 
%language to connect dynamics and computation. Specifically gradient 
%descent along energy landscape provides a framework for understanding 
%how computation of an output state (via local energy minima) can come 
%about in a physical system starting from some initial state that 
%corresponds to an initial system lying in the basin of attraction of 
%the final state. This has been successfully employed for understanding 
%neural complexity using the framework of attractor neural networks 
%(cite Hopfield, and Daniel Amit's book), as well as combinatorial 
%optimization (cite Sherrington and Kirkpatrick "Solvable model of a 
%spin-glass", and Kirkpatrick, Gelatt and Vecchi "Optimization by 
%simulated annealing"). 
%
%However, living systems are always far from 
%equilibrium, and so it would be good to have a corresponding model 
%system that similarly allows us to connect dynamics to computation. 
In this paper, we present a framework for uncovering the computational
logic of systems undergoing transitions between multiple non-equilibrium steady states.
Two archetypes of the irreversible
processes that characterize far-from-equilibrium systems
are diffusion and
chemical reactions (such as combustion) occurring in open 
systems~\cite{Sinha2015}. 
Apart from being fundamental natural processes, they
play important roles in biology, for instance, allowing
transportation of molecules via diffusion, and their
synthesis and replication through metabolic reactions~\cite{Phillips2012}.
Reaction-diffusion systems, in which spatio-temporal patterns can emerge
through self-organization~\cite{Turing1952,Gray1994},
%S. Scott, Chemical Chaos (Oxford University Press, Oxford, 1991).
%S. Sinha and S. Sridhar, Patterns in Excitable Media (CRC Press, Boca Raton, FL, 2015).
provide a natural framework for describing information processing
in terms of out-of-equilibrium dynamics.
Indeed, systems embodying these principles, such as oscillating chemical reactions and
slime molds, have been used to implement specific computational algorithms~\cite{Steinbock1995,Epstein2007,Holley2011,Nakagaki2000,Takamatsu2001,Tero2010,Adamatzky2010}. 
%The plasmodium can be viewed as a system of coupled biochemical oscillators~\cite{Takamatsu2001}
Computation in these systems is implemented by the dynamical evolution of the initial (input)
state to the final (output) state.
The general computational logic presented here suggests that 
%all of these could be part of
these algorithms belong to
a much broader paradigm, in which such evolutions can be completely specified by a set of 
transformation rules (logic) that map the set of non-equilibrium steady states of the system to itself.
By altering the perturbations applied to the system, the logic can be changed,
allowing different types of computations to be realized (analogous to programming a computer).
This can potentially address a central problem in biology: how living organisms ``compute'', i.e., process information vital for survival through biochemical means~\cite{Hjelmfelt1991,Tamsir2011}.
As chemical reactions occurring inside living organisms exhibit temporal recurrence across scales, 
we illustrate this paradigm using a generic non-equilibrium system of relaxation oscillators that are coupled diffusively.
The resulting collective dynamics generates a variety of spatiotemporal patterns
depending on kinetic rates and coupling strengths~\cite{Toiya2008,Singh2013,Menon2014}.
%In this paper we have considered a system of diffusively coupled relaxation
%oscillators that are capable of exhibiting a variety of spatiotemporal patterns
%depending on the kinetic parameters and coupling strength~\cite{Singh2013,Menon2014}. 
%Several of these have in fact been observed in chemical systems~\cite{Toiya2008}. 
Unlike previous attempts at implementing computation in a reaction-diffusion framework,
%setting
which typically involve interactions between traveling waves~\cite{Rossler1974,Hjelmfelt1991,Steinbock1996,Adamatzky2019}, we use a temporally invariant,
spatially heterogeneous non-equilibrium steady state. 
Such Spatially
Patterned Oscillator Death (SPOD) states are observed for relatively strong coupling between oscillators~\cite{Singh2013}.
%In this dynamical regime characterized by time-invariant attractors, we show that a potentially
The temporal invariance allows a particularly succinct dynamical description, and
we show that a potentially
infinite number of phase space trajectories resulting from different initial conditions reduce to
a finite number of qualitatively distinct states. This allows
coarse-graining of the state space continuum to a discrete set that can be mapped
to distinct binary sequences. The time-evolution of the system thereby reduces to operations
in which symbolic strings are transformed to one another. In analogy with computation
in a Turing machine~\cite{Hopcroft2001},
here binary strings representing the initial and final states are the input and output, respectively,
while the role of the program is played by the transformation-inducing perturbation, viz., the binary pattern  of stimuli (i.e., on/off) applied to the array of oscillators. 
%plays the role of the program.
%Providing a succinct description of the dynamics 
%Switching between states is induced by applying appropriate perturbation, which by virtue
%of a binary mapping of the stimulus sequence, represents a program in binary code.
%Mention that we have three binary sequences: input (dynamical state 1), output (dynamical state 2) and the program (perturbation) - and how it fits into the Turing paradigm - you can change the transformation by appropriate program which is also fed as a binary sequence
%By perturbing the system through application of appropriate stimuli, 
%that can also be represented as a binary sequence, switching can be induced between specific 
%pairs of states.
%Indeed, the dynamics, which is effectively reduced to operations in which symbolic strings are transformed to one another can be viewed as computation, inherent or natural to the system. More explicitly, initial and final states represent input and output of the computation, respectively, with the perturbation that transforms one to the other playing the role of the operator. Understanding the processes governing the dynamics thus is akin to inferring the computational logic underpinning the transformations.
We generate a comprehensive taxonomy of the states,
investigate their local, as well as, global stabilities, and provide 
a complete catalogue of perturbation-induced transitions between them,
yielding a fundamental template for realizing computation through the 
collective dynamics of non-equilibrium systems.

\begin{figure}[t!]
\begin{tabular}{c}
\subfigure{\includegraphics[width=\linewidth]{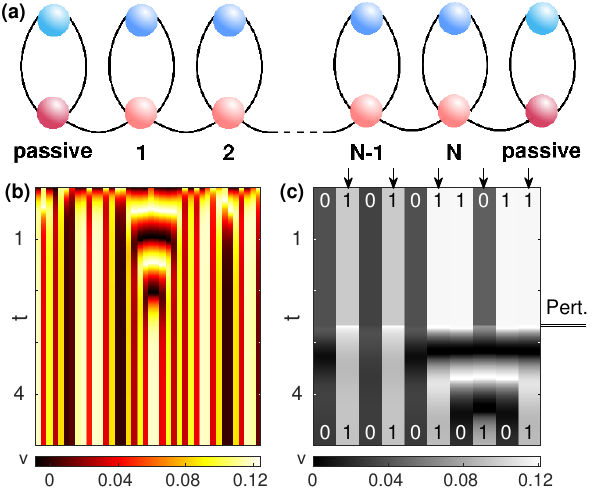}} \\
%{\raisebox{-0.5\height}
\subfigure{\includegraphics[width=5.4cm]{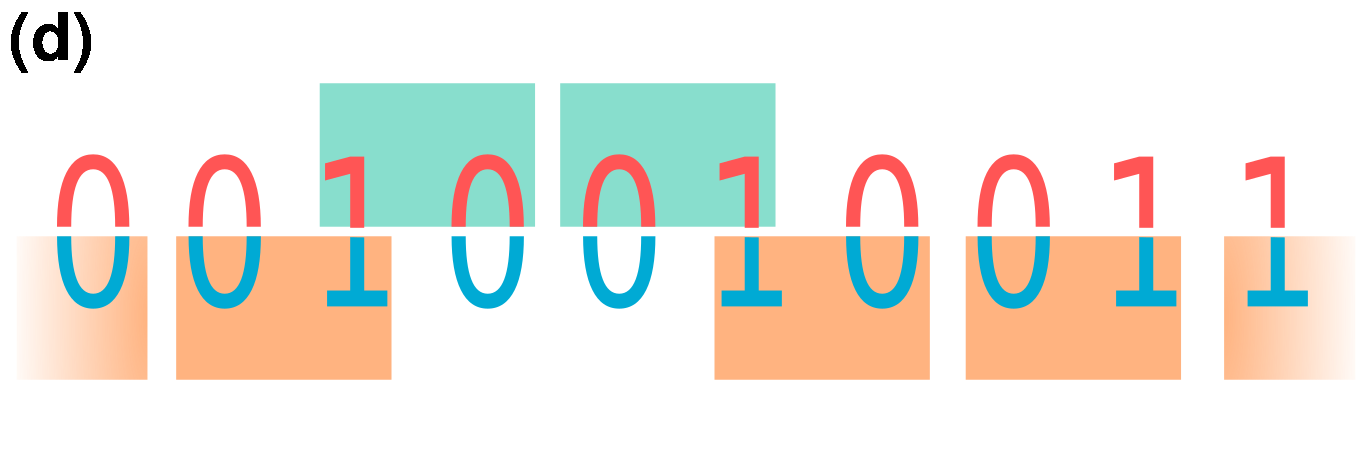}}
\subfigure{\includegraphics[width=3cm]{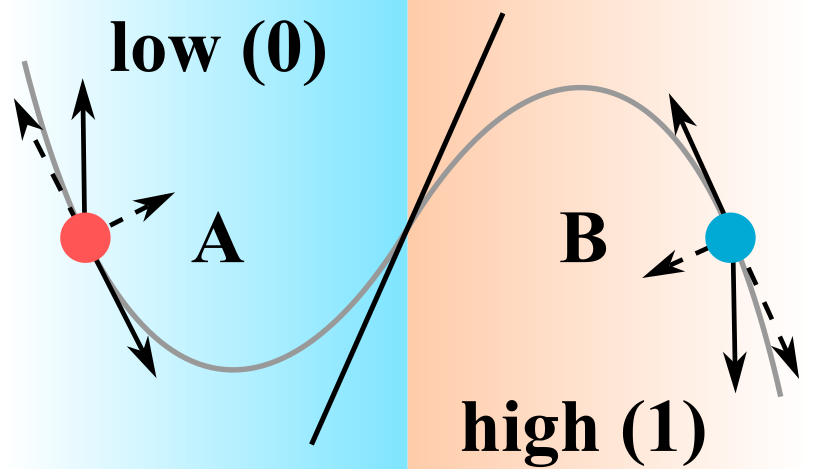}}
\end{tabular}
\caption{
(a) Schematic diagram of a 1-dimensional array of $N$ relaxation oscillators, 
each comprising an activation $u$ (blue) and an inactivation $v$ (red) 
component, diffusively coupled via $v$.
%Nearest neighbors on a $1$-D array are diffusively coupled
%via the inactivation component with strength $D_v$.
%The oscillators are assumed to be identical with passive elements at 
%the boundaries. %MOVE TO TEXT%
(b) Spatiotemporal evolution of $v$ for $N=40$ showing
%The pseudocolor plot of the inactivation variable $v$ shows the 
convergence from a random initial state to a steady state, 
corresponding to a spatially patterned oscillator death (SPOD) configuration. 
Time is normalized by the 
period of an uncoupled oscillator. 
The parameter values are $\alpha=0.139$, $\epsilon = 0.001$, $k=0.6$, 
$b=0.16$ and $D_v=2 \times 10^{-3}$.
(c) Local perturbations
%in which chosen sites are stimulated, can 
induce transitions between the steady states, that are
viewed as binary strings by mapping the low and high values of $v$ to $0$ and $1$, 
respectively.
The outcome of perturbing sites $2, 4, 6, 8$ and $10$, indicated by arrows,
is shown for $N=10$ with signal intensity $A = 8 \times 10^{-4}$ applied over a 
duration $\tau = 3.7 \times 10^{-2}$ (``Pert.''). 
(d)~Each binary string representing a SPOD configuration can be tiled using
pairs of adjacent oscillators in dissimilar states, 
viz., low($0$)-high($1$) or high($1$)-low($0$).
%A representative binary string corresponding to a SPOD state 
%for a system of coupled oscillators on a ring can be tiled using pairs 
%of adjacent dissimilar states (viz, `0-1' or `1-0'). 
There are two reading frames (left, shown using distinctly colored tiles) 
displaced relative to each other by one site, 
which begin at even or odd sites respectively.
Each ``bit'' lies within a tile in at least one 
reference frame, implying that every oscillator in a SPOD state 
is part of a dissimilar ($0$-$1$ or $1$-$0$) pair. 
%for the coupling strength used.
%For a SPOD state to be stable for the coupling strength used, 
%the string should be completely tiled, i.e., each site should belong 
%to a dissimilar pair in one or both reading frames.
%Note that in the above example, each site belongs to at least one 
%dissimilar pair with two sites occurring in overlapping tiles 
%(i.e., they appear in both reading frames) indicating that the 
%state is stable. 
The stability of such dissimilar pairs is explained by the
schematic phase space diagram (right).
%The schematic phase space diagram (right) explains the stability of 
%a pair of adjacent oscillators in dissimilar states. These oscillators,
Oscillators A and B, located on the {\em low} and {\em high} branches of the
cubic nullcline respectively, 
are arrested in a heterogeneous oscillator death configuration.
This results from the balance between components (broken arrows) 
of the opposing forces of coupling and intrinsic kinetics (solid arrows).
%The stability of a dissimilar pair of adjacent sites can be understood 
%from the schematic phase space diagram (right) showing how a pair 
%of coupled oscillators A and B (located on opposite branches of the 
%nullclines) can be arrested in a heterogenic oscillator death configuration. 
%This can be understood from a force balance argument with the 
%opposing forces acting on them, viz. one due to coupling and the other 
%resulting form intrinsic kinetics represented by solid vectors. 
%The components of the coupling force are depicted by broken vectors.  
%(d) Schematic (left) showing a sample binary string `0010010011' which is tiled using high-low pairs (either `0-1' or '1-0'). The two reading frames - one starting from an odd numbered site (green tiles) and the other starting from an even numbered site (orange tiles) tile the string completely. Note that the string is considered circular. There are two sites which appear in overlapping tiles. Schematic (right) showing the nullclines of the FHN system. Two adjacent oscillators A and B are on opposite branches of the cubic nullcline. The two forces acting on either of them - one due to the coupling and the other due to the intrinsic kinetics are depicted by solid vectors. The components of the coupling force is depicted by dashed vectors.
}
\label{Fig1}
\end{figure}

To describe the dynamics of each node in the system of coupled 
relaxation oscillators,
we have used a generic model for such oscillators, viz., the
FitzHugh-Nagumo (FHN) equations.
This model comprises a fast activation variable $u$ and a slow 
inactivation variable $v$ evolving as
\begin{align}
\begin{split}
\dot{u} = f(u,v) & = u(1-u)(u-\alpha) - v\,, \\
\dot{v} = g(u,v) & = \epsilon (k u - v -b )\,,
\end{split}
\label{Model1}
\end{align}
where the parameters $\alpha (=0.139)$ and $k (=0.6)$ describe the kinetics, 
$\epsilon (=0.001)$ governs the recovery rate and 
$b (=0.16)$ characterizes the asymmetry of the limit cycle dynamics. 
%We have verified that small variations around the values used do not affect the oscillatory dynamics.
The simplest coupling topology that 
allows the system to be used for
%provides a framework amenable for
computation is that of a one-dimensional ($1$-D) array, as
each distinct state can be mapped to a symbolic sequence by using appropriate
thresholds.
%In order that the collective dynamics of such oscillators can be used
%to implement computation, the simplest coupling topology that one can
%consider is that of a one-dimensional chain. 
%These sequences can be turned into binary strings on using appropriate 
%thresholds, that can
%function as input and output of the computation process.
%This allows us to view the dynamical evolution of the system between two 
%states as a computation that transforms an input (initial state)
%to an output (final state).
In this paper we have investigated a $1$-D system of
$N$ coupled oscillators with different boundary conditions,
corresponding to a chain and a ring.
Following Ref.~\cite{Singh2013}, we consider the oscillators to be coupled
by diffusion exclusively via the inactivation variable $v$.
Such a coupling captures the essence of the physical process underlying
experiments in microfluidic
devices~\cite{Toiya2008}, where the beads containing the oscillating chemical reactants
are suspended in an inert medium that allows only the inhibitory
species to diffuse.
%
%in experiments on oscillating chemical reactions in microfluidic
%devices, the inert medium in which the beads containing the reactants are
%suspended allow only the inhibitory chemical species to diffuse.
%As we demonstrate later in the paper the dynamics of an array of oscillators
%coupled diffusively can be mapped to binary strings. Thus for our current 
%study we investigate a 1-dimensional with different boundary conditions.
%
%Following the approach of Ref~\cite{Singh2013}, we assume that the oscillators are diffusively coupled through the inactivation variable alone. This is analogous to the experimental case of beads containing reactive solution being suspended in a chemically inert medium that allows passage of only the inhibitory reactant \cite{Toiya2008}. 
A linear chain with 
passive elements at the boundary [Fig.~\ref{Fig1}~(a)] approximates the
geometry of such experimental systems, taking into account the
presence of the inert medium~\cite{Singh2013}.
%By inclusion of non-reactive passive elements at each end of the linear chain, the boundary conditions take the inert medium into account. The inert medium between the oscillators is not considered explicitly. 
The dynamics of the resultant system is described by: 
\begin{equation}
\dot{u}_i =  f (u_i, v_i)\,,
\dot{v}_i =  g(u_i, v_i) + D_v (v_{i-1} + v_{i+1} - 2 v_i) + A_i\,,
\label{Model2}
\end{equation}
%\begin{align}
%\begin{split}
%\dot{u}_i & =  f (u_i, v_i)\,, \\
%\dot{v}_i & =  g(u_i, v_i) + D_v (v_{i-1} + v_{i+1} - 2 v_i)\,, \\
%\dot{v}_0 & =  D_v (v_1 -v_0)\,, \\
%\dot{v}_{N+1} & = D_v (v_N - v_{N+1}) \,,
%\end{split}
%\end{align}
where $ i= 1,2,...,N $ ($N=10$, unless specified otherwise). 
The boundary conditions are specified by
$\dot{v}_0 =  D_v (v_1 -v_0)\,,$ $\dot{v}_{N+1}  = D_v (v_N - v_{N+1}) \,.$ 
The diffusion coefficient $D_v (= 2 \times 10^{-3})$ denotes the strength of coupling between 
the inactivation variables of neighboring relaxation oscillators~\cite{note3}.

The effect of external stimulation on a node $i$ of the system,
such as optical illumination in the case of light-sensitive
chemical reactions that can initiate oscillatory activity in the
medium~\cite{Gaspar1983}, is incorporated through the term
$A_i$ which is assumed to be constant for the duration of stimulation.
%
%The experimental and theoretical studies of the effect of illumination on light-sensitive chemical reactions that can initiate concentration oscillations in the medium gives us an insight into a perturbation mechanism that could be implemented in our model (Eqn.~\ref{Model2}).We incorporate an additional parameter $A_i$, analogous to the intensity of light stimulus, in the equations governing the dynamics of the inactivation variable of the nodes $i$ that are to be perturbed for a duration $\tau$.
%\begin{equation}
%\dot{v}_j = g(u_j, v_j) + D_v (v_{j-1} + v_{j+1} - 2 v_j) + A_{j} \, .
%\label{Model3}
%\end{equation}
%
In the simulations reported here, the signal intensity $A_i$ is assumed to be
the same ($=A$)
for all nodes.
Varying $A$ and the stimulation duration $\tau$ allows us to investigate 
the effect of signal strength on the collective dynamics. 
%potentially providing insights on perturbation mechanisms 
%in experimental realizations of our model system.
%~\cite{note4}.
A stimulation protocol is defined by the choice of nodes $i$ (=$1,\ldots,N$)
for which $A_i \neq 0$ 
%and can be represented by a binary string of length $N$, which results in 
corresponding to a total of
$2^N - 1$ possible protocols.
%excluding the string 0000 . . . 00 which corresponds to no
%stimulation
In principle, these protocols can implement logic gates. Indeed, we explicitly show that
a conservative logic gate~\cite{Fredkin1982} that is capable of universal computation,
can be realized using such protocols~\cite{SI}. 
We have verified that the results reported in this paper are robust with 
respect to different choices of $N$, as well as, small changes in $b, \alpha, k, \epsilon$, 
$D_v$ and $A$.

%\section{Results}

Fig.~\ref{Fig1}~(b) shows the system~(\ref{Model2}) converging to a 
characteristic SPOD state, marked by a temporally invariant pattern
of high and low values for $u_i$, as well as, $v_i$ ($i=1, \ldots, N$). 
Note that, a low (high) value of $u$ implies a low (high) value of $v$
for each element, and vice versa. 
%Furthermore, the high and low values can be distinguished unambiguously.
For each SPOD state, all elements that are low
have exactly the same numerical value for $u$ (and $v$),
as do all elements that are high.
The configuration can hence be mapped to a binary sequence
using an appropriate threshold.
%In the $2N$-dimensional phase space of the system, each SPOD state
%corresponds to two clearly delineated points (corresponding to
%the high and low values).
%In order to map the SPOD pattern from a continuum representation 
%(in terms of $u$ and $v$ values) to a binary string, we specify 
%thresholds for $u$ and $v$ to determine which of the nodes are considered 
%to be in a `high($1$)' or a `low($0$)' state. These threshold values 
%This threshold is chosen by examining the bimodal
As the steady-state probability distributions of $u,v$ (estimated from $10^4$ 
realizations) have a clearly bimodal form, we choose a threshold 
(viz., $u=0.4, v=0.06$) from the region between the peaks,
our results being robust with respect to this choice.
%By examining the bimodal steady-state probability distributions of $u,v$ (estimated from $10^4$ realizations) 
%%and selecting a value 
%%from the region 
%the threshold is chosen from between the two peaks where the probability is essentially zero.
%For the parameter values used here, we have chosen $u=0.4, v=0.06$ as the threshold although the results are robust with respect to small changes in these values. 
To allow an arbitrary binary string to be used
as the initial state of the system [e.g., as in Fig.~\ref{Fig1}~(c)],
we map the entries ($0,1$) of the string to the 
$u,v$ values that occur at the two peaks of the distributions
(viz., $0: u=-0.1, v=0.03$ and $1: u=0.85, v=0.093$). 
%In addition, we use the mode 
%values of these distributions (viz., $u=0.85$, $v=0.093$ for the high state 
%and $u=-0.1$, $v=0.03$ for the low state,which are the 
%exact numerical values of minimum and maximum of $u$ and $v$ corresponding 
%to the most frequently occurring state) to transform a binary string 
%to a continuum representation for the purpose of simulation.

The system exhibits multistability, i.e., for a given choice of parameter 
values, it can converge
to any one of multiple possible SPOD states, each having a corresponding
binary representation.
Fig.~\ref{Fig1}~(c) shows an example of how one such binary string
can be transformed
to another through an external perturbation - a property that is vital
for the system to be used for computation.
While, in principle, a system of $N$ elements should
be able to represent $2^N$ binary strings, 
only a small fraction of these can actually be obtained.
For example, we observe only $68$ distinct strings for a chain of 
$N=10$ oscillators, and $122$ distinct strings for a ring of the same size 
(which reduces to $14$ on considering the cyclic permutations to be 
identical, taking into account the periodic boundary condition).
%We can verify this observation analytically by a linear stability analysis of our model system. We find that the Jacobian of the linearized system at an equilibrium point having the same $(u,v)$ values for any three consecutive oscillators has a positive trace and hence would be an unstable node. (Refer Supporting Information for more details). 
This restriction arises from the physical mechanism by which SPOD states 
are generated,
and can be understood by considering a pair of
coupled oscillators.
When the two oscillators are in opposite branches [Fig.~\ref{Fig1}~(d), right], 
the forces arising from diffusion and intrinsic kinetics acting on the oscillators are balanced under strong coupling, resulting in either a 
$0$-$1$ or a $1$-$0$ state~\cite{Singh2013}. 
%When one considers all SPOD states that can be obtained for a given $N$, 
%starting with random initial conditions, it is observed that they 
%can be represented by a small set of strings. 
For a $1$-D array of $N$ coupled oscillators, we observe that
the SPOD states contain at most two consecutive 0s or 1s, hence
limiting the number of observable binary strings.
%for the chosen values of $b$ and $D_v$.
%~\cite{note1}. 
%we observe that these attainable strings contain 
%a maximum of $k$ $0$s or $1$s, where $k$ is the greatest integer less 
%than or equal to $N/3$. In addition, these strings are seen to contain 
%at most two adjacent $0$s or $1$s. That is, strings containing 
%contiguous sequences of 0s or 1s of length greater than two are not 
%observed to be stable. 

The instability of sequences containing long ($>2$) contiguous blocks of
0s or 1s, for the range of $D_v$ we use, arises because
%when three (or more) adjacent nodes 
%(say, $(i-1)^{th},i^{th}$ and $(i+1)^{th}$ nodes) in the same state, 
the diffusion term for the interior node(s) in such a block is negligible
(thereby disrupting the balance of forces 
responsible for the arrest of oscillations). 
Thus, these nodes effectively become uncoupled and hence oscillate,
%the evolution equation for these nodes
%reduces to the form in Eqn.~\ref{Model1}, implying that they will try to
%oscillate and hence 
destabilizing any time-invariant states
containing such blocks~\cite{SI}. 
%(see Supplementary Information).
This also implies that SPOD states that begin or end with
`00' or `11' will not be observed in a chain with passive elements at its ends~\cite{SI}. 
%As seen from Fig.~\ref{Fig2}~(a),
%the observed SPOD states for a chain of $N=10$ oscillator are constrained
%by this restriction. 
We note also that the inversion of an
observed SPOD state, obtained
by exchanging $0$s with $1$s and vice versa, yields another allowed SPOD state.
The stability of two coupled oscillators arrested in a low and a high
activity state suggests that if such pairs can completely tile the
entire sequence represented by the array [Fig.~\ref{Fig1}~(d), left],
the pattern will be stable.
%If stable low-high node pairs can be used 
%to completely tile the entire sequence [Fig.~\ref{Fig1}~(d), left], 
%the pattern will be stable. 
Indeed, we observe that all SPOD states observed in system~(\ref{Model2})
can be viewed as a sequence of stable pairs of adjacent oscillators, each
pair comprising one oscillator
arrested in a high and the other in a low activity state (i.e., `01' or
`10').
%%%%%%%%%%%%%%PUT IN SI%%%%%%%%%%%%%%%
%Note that when we use such oscillator pairs to tile the chain without any
%overlaps, 
%%(using $N/2$ pairs for any chain of length $N$ that can be completely tiled), 
%there are two possible reading frames, starting 
%from odd and even numbered sites, respectively. 
%The most frequently occurring configurations, viz., those comprising
%a strictly alternating sequence of 0s and 1s,
%can be completely tiled by $N/2$ 
%contiguous stable pairs in either of the reading frames. 
%In contrast, sequences containing
%adjacent oscillators in the same activity state, viz. `00' or `11', 
%can be completely tiled in at most one of the 
%reading frames. For sequences in which neither reading frame leads to a 
%complete tiling, switching between the frames can achieve this 
%[see, e.g., Fig.~\ref{Fig1}~(d), left].
%The number of nodes that can be tiled using stable oscillator pairs 
%in both reading frames (see Fig.~\ref{Fig1}~(d), left) 
%which has a maximum value of $N$
%provides a measure of the stability of the sequence, as discussed later~\cite{SupplementaryInfo}.
%%%%%%%%%%%%%%%%%%%%%%%%%%%%%%%%%%%%%%
%such a defect is encountered in a reading frame, it indicates a switch
%to the other reading frame [Fig.~\ref{Fig1}~(d), left].
%The number of defects present in a string, i.e., the
%number of switches required between reading frames to completely tile it,
%provides a measure of its instability as discussed later.
%If the string cannot be completely 
%tiled under both reading frames, then the state is not a stable one. 
%Furthermore, 
This provides a basis for viewing the 
system as an
anti-ferromagnetic spin chain (spin up/down corresponding to high/low
activity), allowing us to
formulate an effective energy function for the system (see below).

A simple combinatorial argument allows us to obtain the number
of distinct SPOD states appearing in a chain of $N$ oscillators,
$n_{SPOD} (N) = 2 F_{N}$, the $N$-th
term of the Fibonacci sequence~\cite{SI}.
For a ring, a more involved combinatorial argument is required to
determine the distinct number of SPOD states, which in this case 
%ring topology. The number of binary sequences of length $N$ in this case 
is the number of cyclic strings
%i.e., a linear string wrapped at the ends, 
with at most two consecutive 0s and 1s anywhere in the string. 
For a system of size $N$, this number is the coefficient $C(N)$ of the 
generating function 
$f(s) = \sum C(N)s^N = 1 + 2s + 4s^2 + 6s^3 + 6s^4 + O(s^5),$
obtained using the Goulden-Jackson cluster method~\cite{Edlin2000,SI}.
%[see Supplementary Information]. 
Thus, $C(0)=1$, $C(1)=2$, $C(2)=4$ and for $N>2$,
$C(N) = \phi_{+}^{N} + \phi_{-}^{N} + \omega^{-N} + \omega^{N},$ 
where $\phi_{\pm} = \frac{1 \pm \sqrt{5}}{2}$ 
and $\omega = \sqrt[3]{1} $. Taking into account the periodic boundary
conditions
of the ring and using Polya enumeration theory~\cite{Edlin2000}, the number of 
strings unique under cyclic permutations is obtained as
$UC(N) = \frac{1}{N} \sum_{d: d|N} \varphi(d) C'(\frac{N}{d}) \,,$ 
where $C'(1) = 0, C'(2) = 2,$ and $ C'(j) = C(j),$ for $j > 2$. Here
$\varphi (n)$ is the Euler's totient function (or phi function) defined 
as the number of positive integers $\leq~n$ and
relatively prime to $n$.
A numerical evaluation of these formulae show that the
number of distinct SPOD states increases exponentially with 
system size $N$, independent of the connection
topology~\cite{SI}.
%Figure S4
%By employing these formulae, we can see from Fig.~\ref{FigS1} that the 
%number of SPOD states increases exponentially with $N$ in all the 
%three cases, namely, (i) a linear arrangement, (ii) a circular arrangement, 
%and (iii) a circular arrangement where strings unique under circular 
%permutations are considered. 

%These states can be classified based on the number 
%and type of couplets (`00' or `11') in them [Fig.~\ref{FigS0}]. 
%The frequency of their occurrence [Fig.~\ref{Fig2}(a)] indicate that 
%some states are clearly more favored than others. In particular, the 
%most common states are characterized by a regular alternating sequence 
%of $0$s and $1$s, namely, `0101010101' and `1010101010'. 
%When considering $10^4$ random initial conditions, these states are 
%found to occur $1118$ and $1091$ times respectively.   

\begin{figure}[t!]
\includegraphics[width=\linewidth]{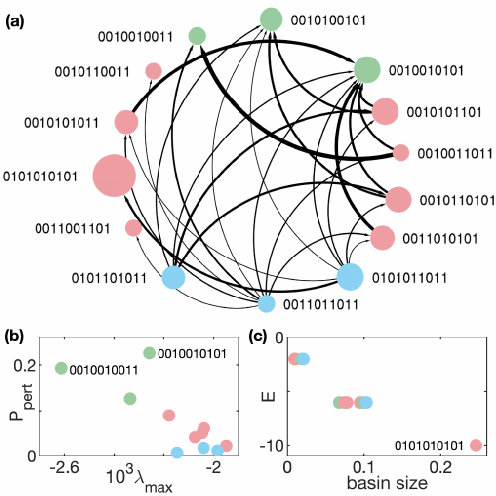}
%\begin{tabular}{c}
%\subfigure{\includegraphics[width=\linewidth]{Fig_2_a.png}} \\
%\subfigure{\includegraphics[width=\linewidth]{Fig_2_b.png}} 
%\end{tabular}
\caption{
%(a) Stylized representation~\cite{WordArt}
%%using a word cloud (powered by \href{wordart.com}{WordArt.com})
%of the attractors of the dynamical system comprising a chain of $N=10$ 
%relaxation oscillators diffusively coupled with $D_v = 2 \times 10^{-3}$.
%The 68 unique SPOD configurations are represented as binary strings,
%where the size of each string is proportional to its basin of attraction.
%These are numerically estimated from their relative frequency of occurrence 
%in $10^4$ realizations using
%random initial conditions. 
%%and no-flux boundary conditions. 
%The strings are colored according to their total number of defects, i.e., 
%pairs of adjacent oscillators in the same activity state: $0$ (red), $1$ (blue),
%$2$ (green), $3$ (brown) and $4$ (black). 
(a) Directed network representing all possible perturbation-induced 
transitions between the 
attractors of a system of $N=10$ oscillators on a ring, coupled diffusively
with $D_v = 2 \times 10^{-3}$.
The $14$ states shown are obtained from the $n_{SPOD}(10)=68$ unique SPOD 
configurations (represented as binary strings) by eliminating all cyclic permutations. 
%having the same number of oscillators [shown in panel (a)]. Note that the
%latter can be obtained through cyclic permutations of the former.
The thickness of a directed link connecting a pair of SPOD configurations 
represents 
the relative number of perturbations that can transform one string to
another. Each of the $1023 (=2^N -1)$ possible perturbation protocols
corresponds to stimulating one or more sites with a signal of 
a specific amplitude $A$ and duration $\tau$
(here $A = 8 \times 10^{-4}$ and $\tau = 5.3 \times 10^{-3}$ units).
Node size is proportional to the size of the basin of attraction
of each configuration, numerically estimated from their relative 
frequency of occurrence in $10^4$ realizations using
random initial conditions. 
%Blue nodes are sources, Green nodes are sinks
(b-c) Stability of the attractors explains
the relative frequency of occurrence of
SPOD configurations. 
%Shown here is a system of $N=10$ relaxation oscillators
%coupled diffusively on a ring with strength $D_v = 2 \times 10^{-3}$
%whose collective dynamics exhibits $14$ attractors.
(b) Local stability of the attractors, 
measured by the largest eigenvalue $\lambda_{\max}$ of
the Jacobian matrix for the corresponding linearized system,
%$2 \times N = 20$ eigenvalues of the Jacobian matrix 
is correlated to the relative frequency of converging to a 
SPOD configuration by perturbing any of the $14$ attractors.
This probability P$_{\rm pert}$ of obtaining a given configuration 
%from the set of attractors 
is computed over all the possible
transformations between states induced by perturbations
%($A = 8 \times 10^{-4}$, $\tau = 10^{-2}$ units)
shown in panel~(a).
%%%%%%%%%
%The amplitude of perturbation is $A=8 \times 10^{-4}$ units and 
%stimulation duration of $ \tau = 10^{-2} $ units. 
(c)~The global stability of each state, quantified in terms
of their basin size, i.e., the probability of occurrence starting from random initial conditions
(estimated from $10^4$ realizations), can be explained by 
an effective energy function $E$
%effective energy function is 
obtained by mapping the system to
an anti-ferromagnetic spin chain (see text).
For all panels, node colors red/blue/green indicate whether a 
configuration has an equal/larger/smaller number of oscillators in high 
activity state (1) than in the low activity state (0), respectively.
Note that, this implies that the blue/green nodes act as sources/sinks, 
respectively.
}
\label{Fig2}
\end{figure}

%We next perform a detailed analysis of the effect of perturbations on the $14$ binary strings that are unique under circular permutations.  As described earlier, a perturbation involves the application of a constant stimulus to a chosen set of nodes for a specified duration, as described in Eqn.~\ref{Model2}. The choice of nodes to be stimulated corresponds to a stimulation protocol which can be represented as a binary string. That is, the nodes to be stimulated are labelled as `1' and are `0' otherwise. Thus, for a ring of size $N$, we have $2^N-1$ stimulation protocols (excluding the string $0000\ldots00$ which corresponds to no stimulation). 

We now investigate the effect of perturbing the SPOD states
that occur in a ring of $N$ elements by using each of the $2^N - 1$ possible
stimulation protocols. Fig.~\ref{Fig2}~(a) shows the perturbation-induced transitions between
strings corresponding to the $14$ distinct states (invariant under cyclic permutations) 
for $N=10$.
%where each string is subject to all the possible $1023$ stimulation protocols.
The number of $0$s present in a string indicates
its relative stability, measured in terms of the number of transitions
that originate from, or terminate at, the string. In particular,
strings with the maximum (minimum) allowed number of $0$s, i.e., $6$ ($4$) 
for $N=10$, 
act as {\em sinks} ({\em sources}, respectively), although 
sources and sinks having $5$ zeros do also occur in this case.
The maximum number $n_{\max} (0)$ of $0$s allowed in a binary string 
of length $N$, subject to the constraint that substrings of contiguous 0s (or 1s) of
length $>2$ are not allowed, is equal to the largest integer $\leq N/3$,
while the minimum number $n_{\min} (0)$ of $0$s allowed is $N - n_{\max} (0)$.
%Pairs of strings that are related by an exchange of the 0s and 1s (i.e.,
%the transformation $ 1 \leftrightarrow 0$) appear as source-sink pairs
%if they have the maximum (minimum) number of allowed 0s, whereas
%they have the same in-degree and out-degree if they have the same number
%of 0s ({\bf what does a source-sink pair mean ?}).
%Thus, the underlying hierarchy 
%(four $\rightarrow$ five $\rightarrow$ six) can be interpreted as 
%transitions from strings of lower to higher stability. The states that 
%are mirror images of each other (that is, which are related by a 
%transformation $ 1 \leftrightarrow 0$) also show interesting behavior. 
%Namely, if these strings have the maximum allowable number of 0s or 1s, 
%they act as source-sink pairs. In other situations, these strings exhibit 
%similar intermediate behavior, i.e. they have the same in-degree and 
%out-degree. We have observed that though 
Note that the strings that are most frequently obtained when starting from random initial conditions 
[the largest nodes in Fig.~\ref{Fig2}~(a)] are not necessarily the ones most stable to perturbation. 
This difference between (i) the probability $P_{\rm pert}$ that a particular 
string results from perturbing the allowed states, and (ii) its basin size, i.e., 
the probability of obtaining the string from random initial conditions,
can be understood in terms of the distinction between
local and global stabilities. 
%That is, if the initial state of 
%the system is random, then we are analyzing its global stability and the 
%SPOD states are the attractors in the $2N$-dimensional system. Whereas, 
%if the initial state of the system is one of the SPOD states and we are 
%considering the effect of perturbations on it, then we are looking at its 
%stability locally.
%States having higher $P_{\rm pert}$ have the least number of perturbation-induced transitions occurring from them, while those 
%having the least $P_{\rm pert}$ have the highest number
%of transitions away from them. This is related to how likely the system will remain
%in a state following perturbation - akin to local stability.

The local stability of a string can be quantified in terms of the decay rate of perturbations to
the corresponding SPOD state, which is related to the largest eigenvalue $\lambda_{\max}$ of the 
Jacobian obtained by linearizing Eqn.~(\ref{Model2}) around this state.
%The local stability of a string can be quantified by measuring how rapidly
%a small perturbation around the equilibrium underlying the associated 
%SPOD state decays with time. This is captured in terms of the maximum
%eigenvalue $\lambda_{\max}$ of the Jacobian matrix of the 
%system (\ref{Model2}) linearized about the equilibrium.
As seen from Fig.~\ref{Fig2}~(b), strings that have the highest $P_{\rm pert}$ (and hence are sinks),
are associated with the most negative values of $\lambda_{\max}$, implying that they
have much higher local stability compared to sources, i.e., strings having the lowest 
$P_{\rm pert}$.
%are more stable against perturbations than the strings corresponding to
%sources, which are seen
%to have the least negative values of $\lambda_{\max}$.
%The more negative the maximum eigenvalue of a SPOD state, the more stable it is to perturbation. Hence, the sinks have the most negative maximum eigenvalues, while the sources have the least negative maximum eigenvalues as shown in Fig.~\ref{Fig6}~(a).
%
The global stability of each string can be defined in terms of an
energy function using the mapping to an anti-ferromagnetic
spin chain. Identifying the $1$ ($0$) in the $i$th node of a string
with $S_i=+1$ or ``up'' ($S_i=-1$ or ``down'', respectively) for the corresponding spin, the
associated energy 
is defined as $E = \sum_{i=1}^{N} S_i S_{i-1}$
%where $S_i \in \{-1,+1\}$ is the spin corresponding to the $i$-th node in the string 
(periodic boundaries imply $S_{0}=S_N$).
Thus, states containing adjacent spins in the same orientation
%Contiguous 0s or 1s in a string contributes to an 
would have a higher energy, decreasing the likelihood that they will emerge
from an arbitrary initial state and hence reducing their basin size [Fig.~\ref{Fig2}~(c)].
%~\cite{note5}.
The use of an energy landscape to describe the system dynamics is consistent
with trajectories initiated from each of the states converging 
to absorbing (sink) states, whose existence implies the absence of 
periodic cycles.
To conclude, the results presented here 
%Our theoretical results 
point to a new computing
paradigm involving the global nonlinear response of a continuous medium 
to local perturbations, which is coordinated through diffusive transport.
Our results could be verified in an experimental set-up with
photoperturbations applied to
arrays of light-sensitive chemical droplets
in microfluidic devices~\cite{Toiya2008,Gaspar1983,Delgado2011}.
As our model is generic,
%with connections to many real world systems, 
it may also be possible to execute such a computation scheme in other
experimental systems involving coupled relaxation oscillators~\cite{Gambuzza14}, including
%including the light-sensitive chemical 
%reactions in mucrofluidic devices~\cite{Toiya2008} described earlier, 
%as well as, 
engineered bioelectric tissues~\cite{McNamara2016}, phototactic
{\em Physarum} plasmodium~\cite{Nakagaki99} and 
insulator-metal state transition devices~\cite{Parihar17}. 
%While the work reported here can be viewed as belonging to the domain
%of reaction-diffusion computation~\cite{Adamatzky2005}, 
%%one of the mechanisms through which nature implements computing~\cite{book},
%we would like to point out a crucial difference. 
Considering strong inhibition between neighboring units allows us to use time-invariant
patterns as the fundamental states in computation involving reaction-diffusion, instead of relying on
interactions between traveling waves~\cite{Adamatzky2019}.
%How our results can be interpreted as chemical logic. 
As there are many biological processes which involve lateral inhibition
between elements having rhythmic activity 
that can be modeled as relaxation oscillators~\cite{Janaki2019}, 
the framework we present here may offer
insights into how computation may be implemented in living 
systems~\cite{Brabazon2015}.

%\section{Acknowledgements}
\begin{acknowledgments}
We would like to thank Alfred A. Aureate and Subinay Dasgupta for 
helpful discussions. 
The simulations required for this work were supported 
by IMSc High Performance Computing 
facility (hpc.imsc.res.in) [Nandadevi and Satpura], which is partially 
funded by the Department of Science \& Technology, Government of India. 
This research has been supported in part by the IMSc Complex Systems
Project (12th Plan), and the Center of Excellence in Complex
Systems and Data Science, both funded by the Department of Atomic Energy, Government of India.
\end{acknowledgments}

%========================================================%
\clearpage
%=====================================================%
\onecolumngrid

\setcounter{figure}{0}
\setcounter{page}{1}
\renewcommand\thefigure{S\arabic{figure}}
\renewcommand\thetable{S\arabic{table}}

\vspace{1cm}
\thispagestyle{empty}
\begin{center}
\textbf{\large{SUPPLEMENTARY INFORMATION}}\\

\vspace{0.5cm}
\textbf{\large{Lateral inhibition in relaxation oscillators provides a basis for computation}}\\
\vspace{0.5cm}
\textbf{A. Parveena Shamim, Shakti N. Menon and Sitabhra Sinha}
\end{center}

\vspace{1cm}
\section*{List of Supplementary Figures}
\begin{enumerate}
\item Fig S1: Stylized representation
of the attractors of the dynamical system comprising a chain of $N=10$ 
relaxation oscillators diffusively coupled with $D_v = 2 \times 10^{-3}$.
\item Fig S2: Statistics of the binary strings corresponding to the
attractors for the dynamical system comprising $N=10$ linearly coupled relaxation oscillators.
\item Fig S3: Generation procedure for binary strings of length $N$ representing the allowed 
SPOD configurations in a system of $N$ oscillators arranged in a chain topology.
\item Fig S4: Dependence of the number of binary strings on the size of the system $N$ for three distinct cases.
\item Fig S5: Alternative representation of the transitions shown in Fig.~2~(a) of the main text.
\item Fig S6: Networks showing the transitions that occur upon perturbing 
SPOD states in a ring of $N=10$ oscillators.
\item Fig S7: Implementation of a Fredkin gate using a given perturbation protocol.
\end{enumerate}

\vspace{1cm}
\section{Numerical details}
The dynamical equations [Eqn.~(2) of the main text] are solved using a fourth order Runge-Kutta scheme 
with time step $\delta t = 0.1$ dimensionless units. 
Time is expressed in units of the period of an uncoupled 
oscillating element.
%In this unit, the stimulus duration $\tau = 4.89 \times 10^{-3}$, unless specified otherwise. 
The initial conditions for each element are randomly sampled from the 
limit cycle of an uncoupled oscillator. 
The simulations reported here use boundary conditions that correspond
to either the chain topology described in the main text [i.e.,
$\dot{v}_0 =  D_v (v_1 -v_0)\,,$ $\dot{v}_{N+1}  = D_v (v_N - v_{N+1}) \,$], or a ring [i.e., $v_0=v_N$, $v_{N+1}=v_1$].
%Unless specified otherwise, we consider the case of $N=10$ oscillators.

The choices of the parameters governing the external stimulation, viz., 
the signal intensity $A$ and the duration $\tau$ for which it is applied,
are constrained to an extent by the requirement that the qualitative nature of the
collective dynamics should not be altered as a result (i.e., a steady state 
configuration  upon perturbation should converge to another steady state,
rather than a spatio-temporally varying pattern).

We note that for much higher values of $D_v$ (e.g., $D_v = 0.015$) 
it is possible to observe contiguous blocks of $0$s or $1$s having length $>2$, 
in particular, $000$ and $111$, but 
the variable $v$ does not have a clear bimodal distribution in this case.
Hence, an unambiguous mapping of the physical state to
a binary string may not be possible for such stronger coupling scenarios.

The absence of SPOD states beginning or ending with `00' or `11' in a
chain having passive elements at its ends,
is a consequence of the boundary condition.
As the the passive elements at the edges will assume
the same activity state (low or high) as that of its neighboring oscillator,
this effectively results in a
block of three consecutive $0$s or $1$s that destabilizes the time-invariant SPOD
state to which it belongs.

\begin{figure}[htbp]
\includegraphics[width=0.85\linewidth]{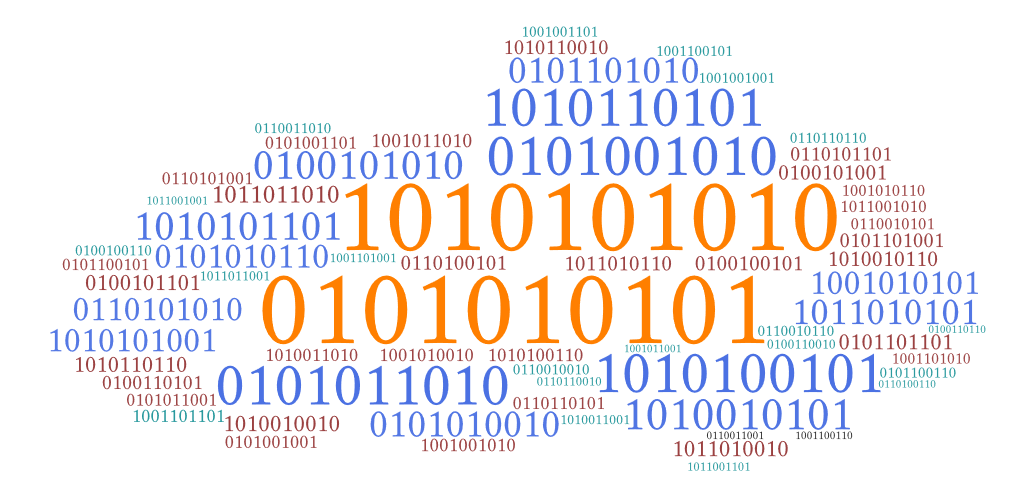}
\caption{Stylized representation
%%using a word cloud (powered by \href{wordart.com}{WordArt.com})
of the attractors of the dynamical system comprising a chain of $N=10$ 
relaxation oscillators diffusively coupled with $D_v = 2 \times 10^{-3}$.
The 68 unique SPOD configurations are represented as binary strings,
where the size of each string is proportional to its basin of attraction.
These are numerically estimated from their relative frequency of occurrence 
in $10^4$ realizations using
random initial conditions. 
%%and no-flux boundary conditions. 
The strings are colored according to their total number of defects, i.e., 
pairs of adjacent oscillators in the same activity state: $0$ (red), $1$ (blue),
$2$ (green), $3$ (brown) and $4$ (black). The visualization has been generated using
the online tool provided by \href{wordart.com}{WordArt.com}.
}
\label{Figwordle}
\end{figure}

Fig.~\ref{Figwordle} shows all the 68 distinct SPOD configurations (represented as binary
strings, see main text for details about the corresponding mapping) obtained as attractors of the dynamics
for a chain of $N=10$ oscillators, out of the $2^{10}$ possible binary states that the
system can adopt during its time-evolution. The statistics for the attractors, including estimates
of the sizes of their basins of attraction, are obtained by performing $10^4$ different realizations 
of the system dynamics using randomly chosen initial configurations.

\begin{figure}[h]
\includegraphics[scale=0.45]{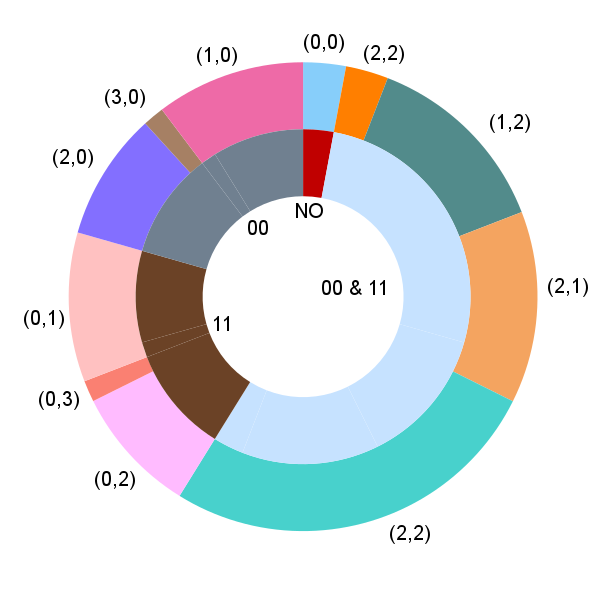}
\caption{Statistics of the binary strings corresponding to the
attractors for the dynamical system comprising $N=10$ linearly coupled relaxation oscillators.
The $68$ distinct attractors (each representing a SPOD configuration) of the system
are characterized in terms of the number of defects (i.e., adjacent pairs of elements in 
identical low or high activity states) that result in the string deviating from a strictly 
alternating sequence of 0s and 1s (i.e., `$0101\ldots01$' or `$1010\ldots10$'). 
In each wheel, the sizes of the various colored segments indicate the relative
fraction of the corresponding string type represented by the segment.
The inner wheel provides a gross classification between string fractions with NO defects, those
having defects strictly of the type `$00$' (i.e., adjacent elements in low activity states) or `$11$'
(i.e., adjacent elements in high activity states), and those having defects of both types `$00$' and
`$11$'. The outer wheel shows a finer subdivision of the segments shown in the inner wheel,
and provides information about the relative fraction of strings that have a specific 
number of `$00$' and `$11$' defects. 
The ordered pair ($X,Y$) corresponding to each segment indicates that the string has
$X$ pairs of adjacent elements in low activity states and $Y$ pairs in high activity states.}
\label{FigS0}
\end{figure}

\section{Reading Frames}
In Fig.~1~(d, left) in the main text we have shown the tiling of a chain of $N$ oscillators by stable pairs of adjacent oscillators arrested in low and high activity states, respectively.
Note that when we use such oscillator pairs to tile the chain without any
overlaps, 
(using $N/2$ pairs for any chain of length $N$ that can be completely tiled), 
there are two possible reading frames, starting 
from odd and even numbered sites, respectively. 
The most frequently occurring configurations, viz., those comprising
a strictly alternating sequence of 0s and 1s (see, e.g., Fig.~\ref{Figwordle} for $N=10$),
can be completely tiled by $N/2$ 
contiguous stable pairs in either of the reading frames. 
In contrast, sequences containing
adjacent oscillators in the same activity state, viz. `00' or `11', 
can be completely tiled in at most one of the 
reading frames (see Fig.~\ref{FigS0} for a graphical overview of the statistics
of strings with such `defects'). For sequences in which neither reading frame leads to a 
complete tiling, switching between the frames can achieve this 
[see, e.g., Fig.~1~(d), left, in main text].
The number of nodes that can be tiled using stable oscillator pairs 
in both reading frames (which has a maximum value of $N$)
provides a measure of the stability of the sequence.
In particular, if the string corresponding to a state cannot be completely 
tiled under both reading frames, then the state will not be stable, and hence will not be observed.

\section{Number of binary strings in different connection topologies}
We provide below combinatorial arguments for obtaining the number of distinct
SPOD configurations represented
using binary strings in a 1-dimensional array of $N$ oscillators with different boundary conditions.
\subsection{Chain of $N$ oscillators}
\begin{figure}[h]
\includegraphics[width=0.9\textwidth]{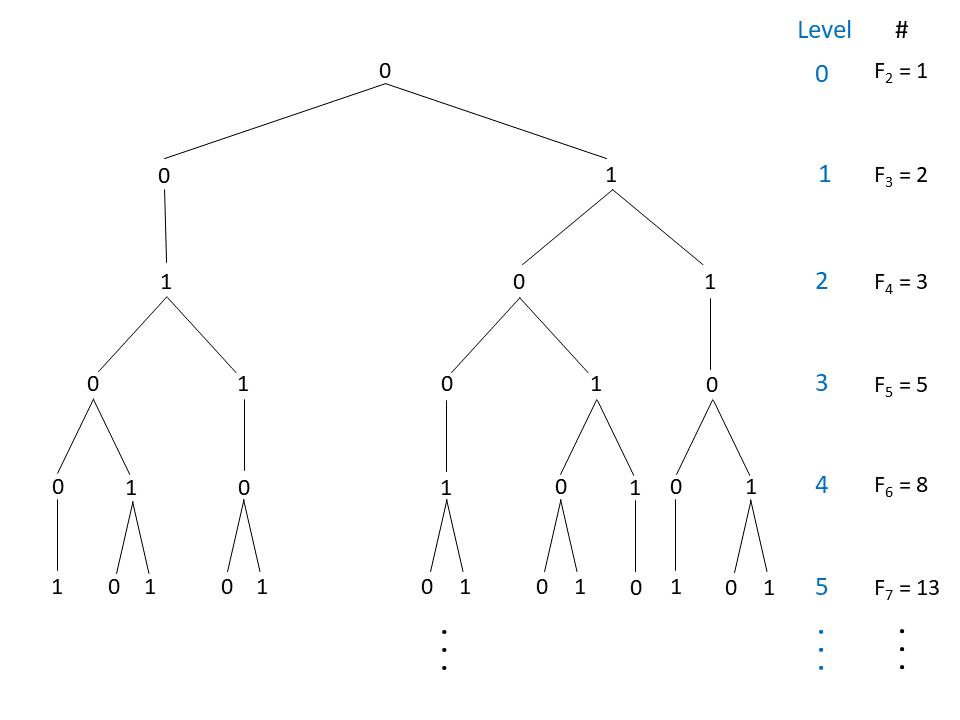}
\caption{
Generation procedure for binary strings of length $N$ representing the allowed 
SPOD configurations in a system of $N$ oscillators arranged in a chain topology.
At each level $L (=N-1)$, the number of sequences allowed subject to the constraint that
there cannot be more than $2$ consecutive $0$s or $1$s is given by $F_{N+1}$, the
$(N+1)$-th entry of the Fibonacci sequence (see text).}
\label{binarytree}
\end{figure}

To calculate the number of SPOD configurations in a chain (i.e., where the oscillators
can be uniquely identified in terms of their position in the sequence in which they occur in the array),
we note that this can be expressed by first finding the number $n_1(N)$ of binary strings of length $N$
which do not contain more than two consecutive $0$s or $1$s (as this is disallowed under
the system dynamics for the parameter values used). We then remove from this set all
strings, say $n_2 (N)$ in number, which have two consecutive $0$s or $1$s at either boundary 
of the chain (as the
boundary conditions prevent the dynamics at the chain ends from generating such patterns).
For a given $N$, the number $n_1(N)$ is simply obtained by counting the strings generated
using a pair of binary trees, one beginning with $0$ and the other with $1$ at its root, and
terminating at level $N-1$, subject to the constraint that the strings of length $N$, 
obtained by following along every 
branch the binary digits occurring at each level from the root (level $0$) to leaf (level $N-1$), should not have more than two consecutive $0$s or $1$s. As can be seen from
Fig.~\ref{binarytree}, the number of such sequences for each of the two binary trees is
given by $F_{N+1} (=F_{N}+F_{N-1})$, i.e., the ($N+1$)-th Fibonacci number ($F_1=1$,
$F_2=1$, $F_3=2$, $F_4 = 3$, \ldots). As the two binary trees are identical with only $0$
and $1$ interchanged, this implies that $n_1(N)=2 F_{N+1}$.
Similar arguments yield $n_2(N)=2 F_{N-1}$. Thus, the total number of 
SPOD states for $N$ oscillators arranged in a chain topology is
$n_{SPOD} (N) = n_1 (N) - n_2 (N) = 2 F_N$.

\subsection{Ring comprising $N$ oscillators}
We now obtain the number of distinct
SPOD configurations in a ring topology, i.e., which are invariant under cyclic permutations.
Using Goulden-Jackson cluster method, the generating function for the number of cyclic strings is: 
\begin{align}
\begin{split}
%\begin{multiline}
f(s) & =  \frac{(1 - 6s^3 + 6s^4 - s^6) }{(1 - s^3) (1 - 2s + s^3)} +  2 \left( Chop_3 \left( \frac{s^2 ( 1 + s) (1+2s)}{1 + s + s^2} \right) \right) \\
%\end{multiline}
& =  1 + 2s + 4s^2 + 6s^3 + 6s^4 + 10s^5 + 20 s^6 + O(s^7) \,, \\
& = \sum C(N) s^N \,,
\end{split}
\label{Formula1}
\end{align}
where $Chop_r ( \sum_{p=0}^{\infty} a_p s^p)  : = \sum_{p=r}^{\infty} a_p s^p $. 
The coefficients of $f(s)$ are in general given by: 
\begin{equation}
C(N) = 
\left\lbrace
\begin{array}{c c}
2 & \mbox{if} \, N=1 \\
4 & \mbox{if} \, N=2 \\
\phi_{+}^N + \phi_{-}^N + \omega^{-N} + \omega^N & N>2
\end{array} \right. \,,
\label{Formaula2}
\end{equation}
where $ \phi_{\pm} = \frac{1 \pm \sqrt{5}}{2}$ and $\omega = \sqrt[3]{1}$. \\
\begin{figure}[h]
%\begin{tabular}{c}
%\subfigure{\includegraphics{PolarFreqBar3_crop.pdf}}\\
%\subfigure{}
\includegraphics{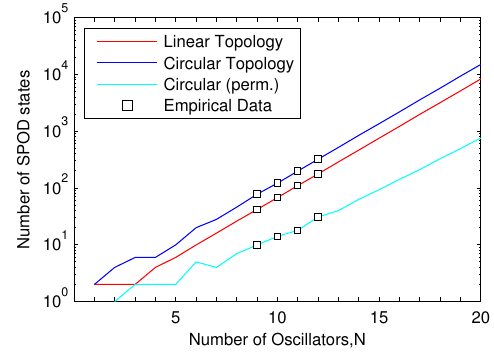}
%\end{tabular}
\caption{
%(a)  Relative frequency of the binary strings seen for the case of $N=10$ oscillators arranged in a ring, consisting of those strings that are unique under circular permutations. The length of the bars represent the frequency of occurrence of each of these strings (out of $10^4$ unique random initial conditions). The bars of same color correspond to those strings which are mirror images of each other ( i.e. $1 \rightarrow 0$ and $0 \rightarrow 1$).
%(b) 
Dependence of the number of binary strings on the size of the system $N$ for three distinct cases: (i) a linear arrangement, (ii) a circular arrangement, (iii) a circular arrangement where only strings unique under circular permutations are considered. The formulae used to determine the number in each case are detailed in the main text.
}
\label{FigS1}
\end{figure}

\noindent
The number of strings unique under cyclic permutations is
\begin{equation}
UC(N) = \frac{1}{N} \sum_{d: d|N} \varphi(d) C'\left(\frac{N}{d}\right) \,, 
\label{Formula3} 
\end{equation}
where 
\begin{align}
\begin{split}
C'(j)& = 
\left\lbrace
\begin{array}{c c}
0 & \mbox{if} \, j=1 \\
2 & \mbox{if} \, j=2 \\
C(j) & j > 2
\end{array} \right. \\
& = \left\lbrace \begin{array}{c c}
0 & \mbox{if} \, j=1 \\
2 & \mbox{if} \, j=2 \\
\phi_{+}^j + \phi_{-}^j + \omega^{-j} + \omega^j & \mbox{if} \, j > 2
\end{array} \right. \, .
\end{split}
\label{Formula4}
\end{align} 
and $\varphi (n)$ is the Euler's totient function (or phi function) defined as
the number of positive integers $\leq~n$ and relatively prime to $n$ .

Fig.~\ref{FigS1} shows for different topologies, viz., the chain and the ring, the nature
of increase in the number of distinct SPOD configurations.

\section{Perturbation-induced transitions}
\begin{figure}[h]
\includegraphics{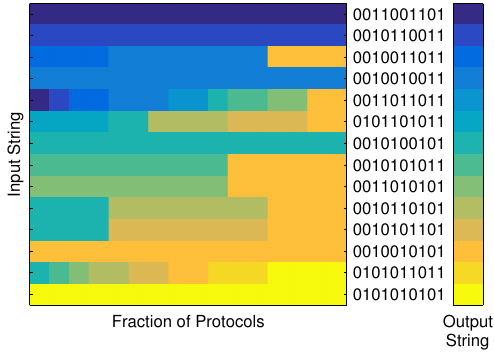}
\caption{
Alternative representation of the transitions shown in Fig.~2~(a) of the main text. The rows correspond to the 14 input stings, the colors denote the output strings and the relative width of the bands in each row indicate the fraction of protocols (out of 1023) that lead to a given output state.}
\label{FigS2}
\end{figure}

As mentioned in the main text, perturbing a SPOD state in an array of $N$ coupled oscillators,
using any one of the $2^N - 1$ possible protocols for stimulating the various distinct possible
combinations of oscillators, results in the system converging to any of the dynamical attractors
(corresponding to distinct SPOD states). Fig.~2~(a) in the main text
represents the entire set of such
perturbation-induced transitions between the 14 possible distinct SPOD states in a system of $N=10$
oscillators arranged in a ring topology (diffusive coupling strength $D_v = 2 \times 10^{-3}$).
While the width of a directed link between two SPOD states $X,Y$ in Fig.~2~(a) have been made proportional 
to the number of perturbations that drive a given input state $X$ to a given output state $Y$,
in Fig.~\ref{FigS2} we present the information about these transitions using an alternative
representation that allows certain information about these transition to be gleaned more conveniently.  
In particular, we immediately note from Fig.~\ref{FigS2} that $6$ of the SPOD states 
(viz., $0011001101$, $0010110011$, $0010010011$, $0010100101$, $0010010101$ and
$0101010101$) are stable
configurations, in that no perturbation applied on such a string can result in a transition to some 
other state. 

The representation also allows us to infer the relative fraction of different perturbations
of various states that converge to a particular output state by looking at the total area of blocks 
having the same color. For instance, Fig.~\ref{FigS2} indicates that the stable state $0010010101$
can be arrived at from many of the other SPOD states through the application of a  large number of possible perturbations. Note that, this is quite distinct from the concept of `basin of attraction'
of a SPOD state $X$,
which asks how many different randomly chosen initial configurations (which need not correspond
to a SPOD state) converge to $X$. Thus, for example, in this case, the state having the largest
basin size is $0101010101$ [Fig.~2~(a)], which has fewer perturbation-induced transitions to it than the
aforementioned SPOD state $0010010101$. This difference can be traced to the distinction
between local and global stability as discussed in the main text.
%
%\begin{figure}[h]
%\includegraphics{Alluvial.png}
%\caption{
%Alternative representation of the transitions in [Fig.~\ref{Fig2}~(a) ]. The strings shown
%in the left correspond to the input states (on which a perturbation is applied), 
%while the ones in the right correspond to the output resulting from a perturbation.
%}
%\label{FigS2}
%\end{figure}

\begin{figure}[h]
%\begin{tabular}{c}
\includegraphics[width=0.48\textwidth]{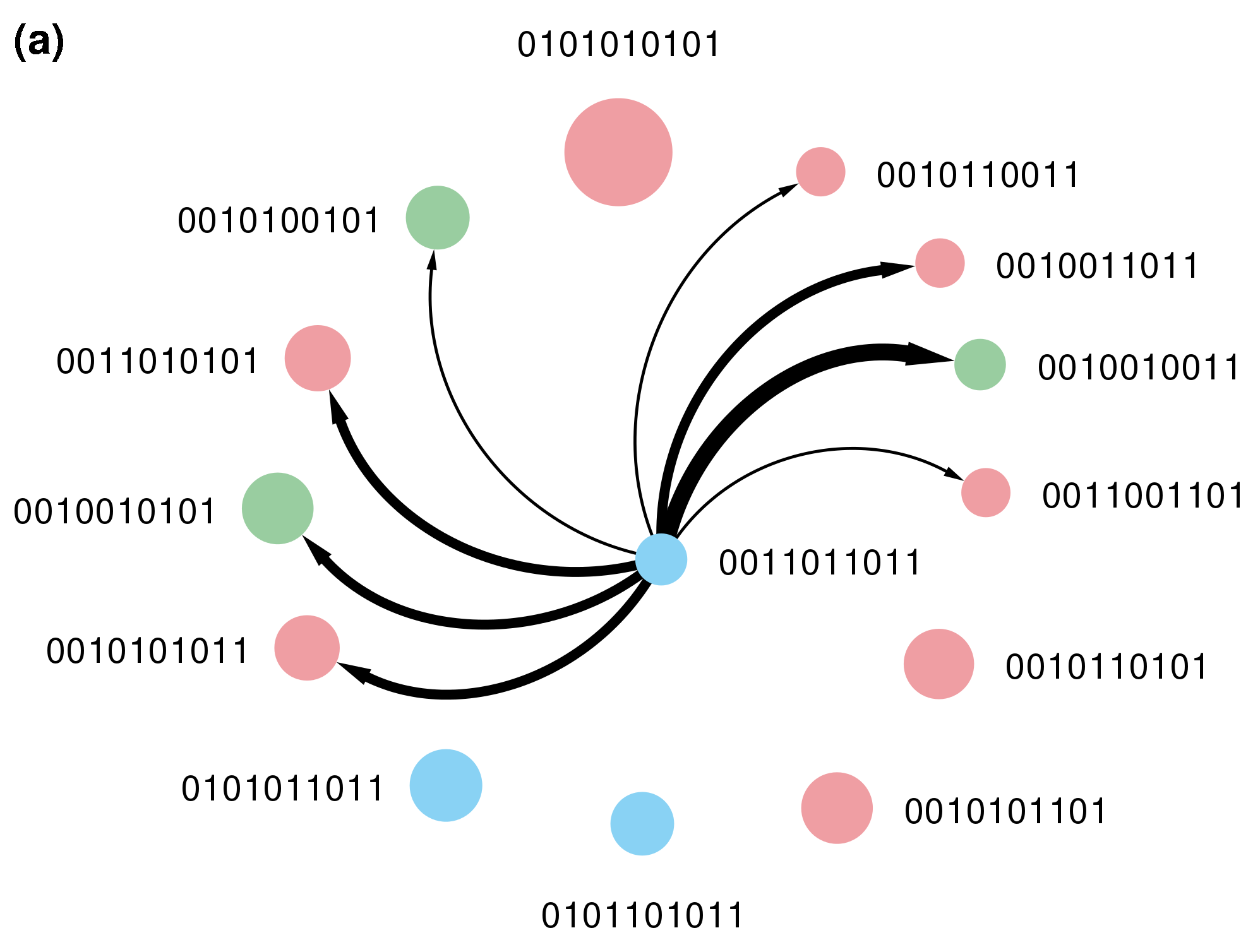}
\includegraphics[width=0.48\textwidth]{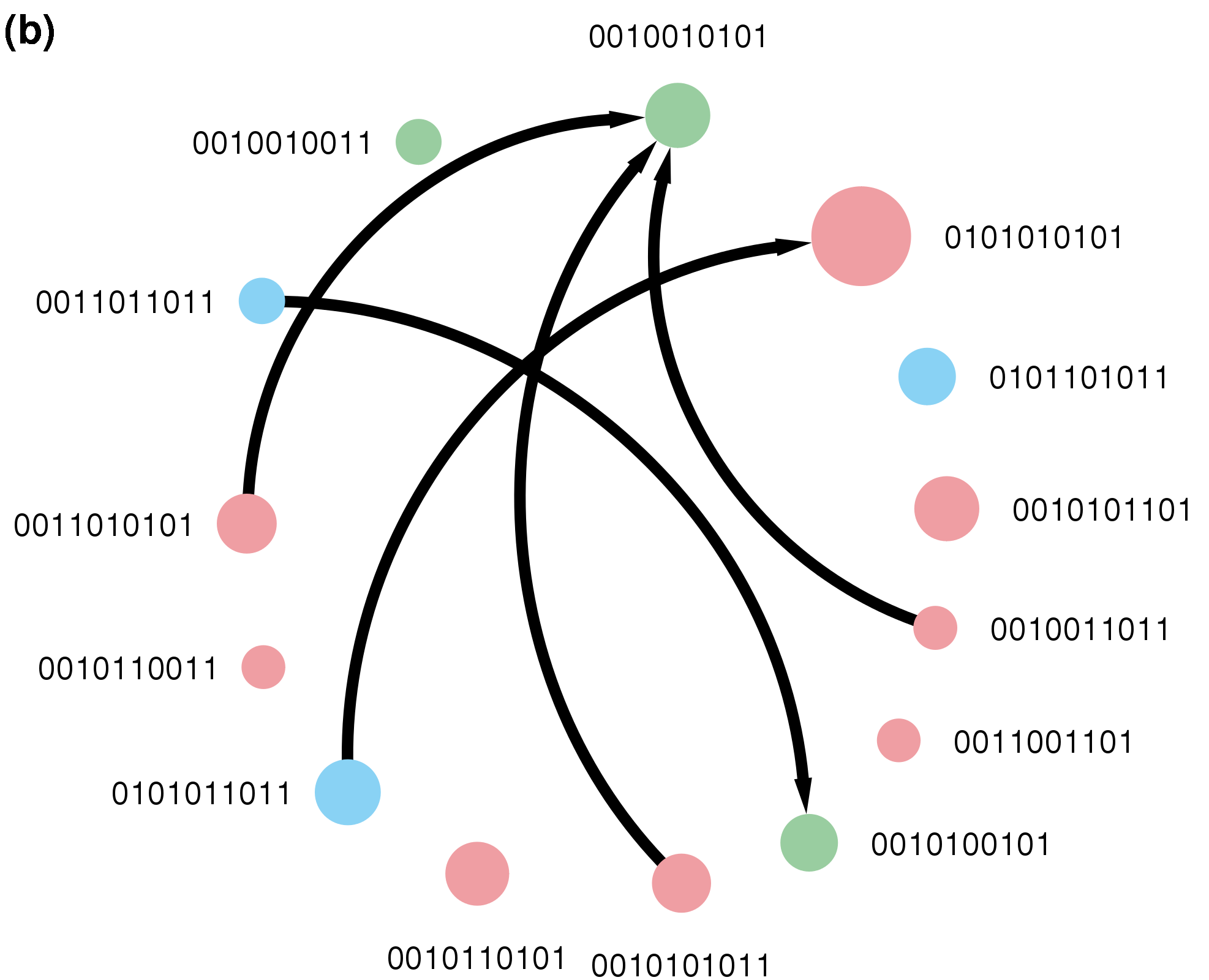}
%\subfigure{\includegraphics{String5.png}} 
%\subfigure{\includegraphics{Prtcl_1024.png}} 
%\end{tabular}
\caption{Networks showing the transitions that occur upon perturbing 
SPOD states in a ring of $N=10$ oscillators. The amplitude and duration
of the perturbation is the same as in Fig.~2 of the main text,
as is the convention used for node color, node size
and edge thickness.
(a) The transitions that can occur from the SPOD state represented
by `0011011011' when it is subject to all possible $1023$ stimulation protocols.
We note that the transitions appear to be governed by the rule that a state
with a certain number of 0s can only transition to one
with a larger number of 0s. This implies a hierarchy among the allowed
SPOD states, viz., a string with four 0s can transition to one with either
five or six 0s, while a string with five 0s can only transition to a string
having six 0s. This is illustrated in panel (b), which displays the
transitions that occur when the 14 distinct SPOD states are subject to the 
protocol in which all sites are stimulated. Even though all nodes are
perturbed in this case, transitions are observed only for a few strings while
the others remain unchanged (note that, the edges are unweighted
%, and indicate the final state reached in each case. Note 
and that self-loops have not been displayed.}
%Networks showing the transitions that occur upon perturbing SPOD states in a ring of $N=10$ oscillators, coupled through the inactivation variable with $D_v = 2 \times 10^{-3}$. A stimulation protocol corresponds to perturbing chosen set of sites among $ N = 1, 2 ... 10$ with a fixed amplitude ($8 \times 10^{-4})$ and duration ($0.125$ units). The states represented as the nodes in the network correspond to those that are unique under circular permutations. That is, an arrow between the original and perturbed states indicate that one of the possible circular permutations of the final state has been reached. The colors of the nodes indicate the number of zeros (blue : 4, red : 5, green : 6) in the corresponding binary representation of the SPOD state, and the relative node size denotes the frequency of occurrence of the state from $10^4$ random initial conditions (with large sizes denoting states that occur more frequently).
\label{FigS3}
\end{figure}

We also note from Fig.~\ref{FigS2} that input states differ in terms of the number of 
distinct output states that can be reached via different perturbations. 
%While for stable states, 
While stable input states each have an unique output state (namely itself)
%there is a unique output state for each input state (namely itself) 
which is
independent of the perturbation applied, other states
could map to a few (e.g., states $0010101011$ and $0011010101$ map to either themselves or to $0010010101$ depending on the perturbation) or many output states. An example of the latter
is the input state $0011011011$, whose perturbation-induced transitions are shown in detail
in Fig.~\ref{FigS3}~(a).

In order to understand how the application of a particular perturbation to 
each of the distinct input states in turn results in transitions to different output states, we have considered in Fig.~\ref{FigS3}~(b)
the protocol in which all oscillators are stimulated
%(i.e., the protocol $1111111111$). 
While $9$ of the states are left unchanged by the perturbation, $3$ of the remaining states
map to $0010010101$ (the state which is the target of the largest number of perturbation-induced
transitions, as we have seen from Fig.~\ref{FigS2}) while one state maps to 
$0101010101$ (the state having the largest basin) and the other maps to $0010100101$.
Note that while, in general, states having the maximum number of $0$s (nodes colored green)
act as sinks and those having the minimum number (nodes colored blue) act as sources,
not all such nodes behave similarly for a given perturbation. 
For instance,  $0101101011$ (a blue colored node) does not map to any other state
under this particular stimulation protocol, while $0010010011$ (a green colored node) 
is arrived at only from itself.

\section{Implementation of a conservative logic gate}
Here we show that an array of coupled oscillators subject to a given perturbation can be used to implement
a conservative logic gate, specifically the Fredkin gate, which is  known to be universal (in the sense that
any logical operation can be realized by suitable combinations of these gates; see Ref.~[44] 
%~\cite{Fredkin1982}
cited in the main text for details). Operations performed by this gate are reversible, such that the 
transformation of an input configuration $I$ to a given output $O$ can be inverted by connecting it to another gate,
creating a cascade. The Fredkin gate $F$ is therefore its own inverse $F^{-1}$, where $F^{-1} ( O ) =  F^{-1} ( F (I)) = I$.

%If we use the ``spin'' representation for configurations as described in the main text,
%viz., identifying the $1$ ($0$) in the $i$th node of a string
%with $S_i=+1$ or ``up'' ($S_i=-1$ or ``down'', respectively),
%we can define an input/output channel $k$ (input:~$I_k$, output:~$O_k$) in terms of a set $\mathcal{C}_k$ comprising an odd number of sites $j$
%whose identities are invariant for a given realization of the gate. The state of a channel
%is determined by a majority rule over the states of the constituent sites,
%such that if $\mathbf{S}^{init}$ and $\mathbf{S}^{final}$ represent the initial
%and final states of the spins representing the sites, then the state of the input channel $k$, $I_k = \operatorname{sgn} (\sum_{j \in C_k} S^{init}_j)$
%and the state of the output channel $k$, $O_k = \operatorname{sgn} (\sum_{j \in C_k} S^{final}_j)$, where $\operatorname{sgn} (x)
%= +1$, if $x>0$, = 0 if $x=0$, and $= - 1$, otherwise. 
\begin{figure}[b]
%\begin{tabular}{c}
\includegraphics[width=0.99\textwidth]{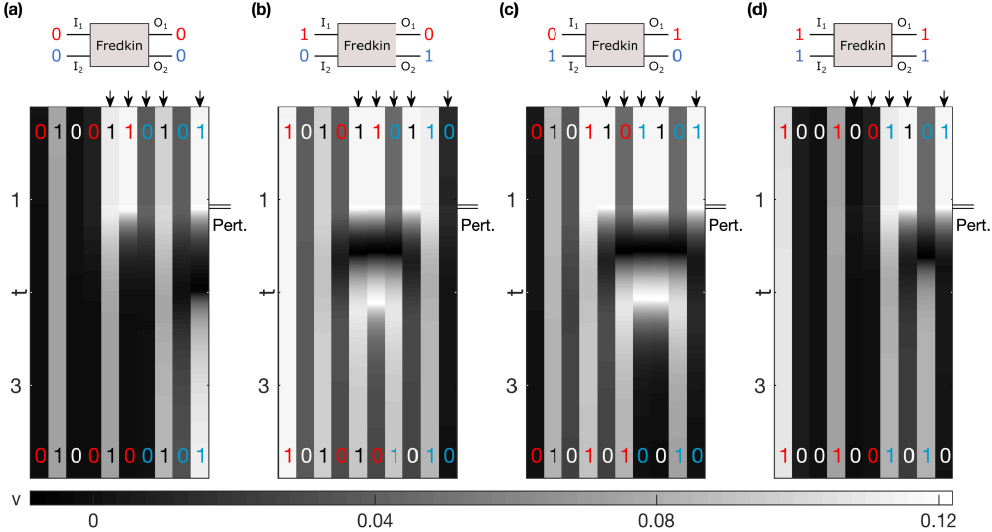}
%\end{tabular}
\caption{Implementation of a Fredkin gate using a given perturbation protocol.
The upper row schematically depicts the various input/output configurations (corresponding to the different rows of the truth table)
of the $2$-input, $2$-output gate when the control line $C=1$, i.e., in the case where the perturbation is applied. Note that, $C=0$ corresponds to the case where no perturbation
is applied, and consequently the SPOD state is unchanged (not shown).  Specifically, panels (a-d) show the transformations
(a) $\{I_1=0,~I_2=0\}\rightarrow\{O_1=0,~O_2=0\}$, (b) $\{I_1=1,~I_2=0\}\rightarrow\{O_1=0,~O_2=1\}$,
(c) $\{I_1=0,~I_2=1\}\rightarrow\{O_1=1,~O_2=0\}$, and (d) $\{I_1=1,~I_2=1\}\rightarrow\{O_1=1,~O_2=1\}$.
This is implemented in a $1$-dimensional array comprising $N (=10)$ oscillators coupled diffusively to their nearest neighbors with
$D_v = 2 \times 10^{-3}$, by subjecting the sites  $5$, $6$, $7$, $8$ and $10$ to a perturbation of amplitude $A = 8 \times 10^{-4}$ 
for a duration $\tau = 10^{-2}$ time units. The boundary conditions for an open chain are implemented by coupling passive cells to
the oscillators at the extremities (i.e., oscillators $1$ and $N$ are each coupled to one oscillating and one passive cell). 
The bottom row shows the transformations realizing each
row of the truth table below the corresponding schematic input/output configuration, as a 
spatio-temporal evolution of different SPOD states (indicated by the binary string superposed on the array) subjected to the specified perturbation (indicated by the arrows) 
and consequently behaving as a Fredkin gate ($C=1$).
As mentioned in the text, each input/output channel comprises a set of three sites, viz., $I_1/O_1$ corresponds to the the sites $1$, $4$ and $6$ (whose binary entries
are colored red), while the channel $I_2/O_2$ comprises sites $7$, $9$ and $10$ (binary entries colored blue). 
The state of the channels are determined by the majority rule (see text).}
\label{Fig_fredkin}
\end{figure}
The Fredkin gate is typically represented as a gate which has $3$ input channels ($C$, $I_1$, $I_2$) and $3$ output channels ($C$, $O_1$, $O_2$), each of which
have binary values ($0$ or $1$). 
The control line $C$ is unchanged by the gate, and we identify its value as indicating either the application of a given perturbation ($C=1$) or not ($C=0$).
The input and output channels are represented as sets comprising an odd number $M$ of sites in the array ($M=3$ for the results shown here).
The value assigned to each channel is obtained by applying the majority rule on the states of the constituent sites.   
Thus, if a channel consists of three sites (i.e., $M=3$) with at least two in the state $1$, then the value assigned to the channel is $1$; else it is assigned $0$.  
A given pair of input and output channels $I_j$, $O_j$ ($j=1,2$) share the same sites, such that the logical operation of the
gate is identified with the state transformation resulting from the perturbation. 

Fig.~\ref{Fig_fredkin} shows a realization of the Fredkin gate using a protocol where, if the perturbation is applied ($C=1$), the sites 
$5$, $6$, $7$, $8$ and $10$ are stimulated with  amplitude $A$ for a duration $\tau$, while $C=0$ corresponds to absence of 
perturbation such that the state of the array remains unchanged (the panels show only the non-trivial case $C=1$).
As can be seen, each of the rows of the Fredkin gate truth table for $C=1$ can be implemented using this perturbation.
We have verified that other perturbation schemes also realize this logic.
\end{document}